\documentclass[sigconf,nonacm]{acmart}

\AtBeginDocument{%
  }

\title{
Developing a Risk Identification Framework\\for Foundation Model Uses
}

\author{David Piorkowski}
\email{David.Piorkowski@ibm.com}
\affiliation{%
  \institution{IBM Research}
  \city{Yorktown Heights}
  \state{New York}
  \country{USA}
}

\author{Michael Hind}
\email{hindm@us.ibm.com}
\affiliation{%
  \institution{IBM Research}
  \city{Yorktown Heights}
  \state{New York}
  \country{USA}
}

\author{John Richards}
\email{ajtr@us.ibm.com}
\affiliation{%
  \institution{IBM Research}
  \city{Yorktown Heights}
  \state{New York}
  \country{USA}
}

\author{Jacquelyn Martino}
\email{jmartino@us.ibm.com}
\affiliation{%
  \institution{IBM Research}
  \city{Yorktown Heights}
  \state{New York}
  \country{USA}
}

\begin{document}

\begin{abstract}
    As foundation models grow in both popularity and capability, researchers have uncovered a variety of ways that the models can pose a risk to the model's owner, user, or others. Despite the efforts of measuring these risks via benchmarks and cataloging them in AI risk taxonomies, there is little guidance for practitioners on how to determine which risks are relevant for a given foundation model use. In this paper, we address this gap and develop requirements and an initial design for a risk identification framework. To do so, we look to prior literature to identify challenges for building a foundation model risk identification framework and adapt ideas from usage governance to synthesize four design requirements. We then demonstrate 
    how a candidate framework can addresses these design requirements and provide a foundation model use example to 
    show how the framework works in practice for a small subset of risks.
\end{abstract}

\maketitle

\section{Introduction} \label{sec:intro}
Due to their generality, foundation models have been used in many situations, such as summarizing text, answering questions, and providing strategies to accomplish tasks. However, with these capabilities comes well-documented risks~\cite{risk-atlas-paper, risk-atlas-url, owasp, mit-risk-repo,nist-rmf}, which are unacceptable harms to individual users, groups, AI model deploying organizations, or society at large.
Thus, the importance of governing artificial intelligence (AI) systems increases as foundation model capabilities continue to improve. 

AI governance is the process of managing these risks throughout the AI development and deployment lifecycle. It 
starts with identifying risks before deployment and includes quantifying and mitigating these risks both before and
after deployment. 
However, recent examples showcase how current governance practices of foundation models remain immature, resulting in critical failures of AI systems~\cite{samsung-data-leak,ny-lawyers-hallucination,air-canada-chatbot,support-agent-hallucination,nyc-chatbot}. 

A key challenge in governing foundation models is the multitude of potential uses of these models. An AI model's use determines what risks are relevant and thus, how it should be governed. This in turn, determines what risk measurements should occur, what risk mitigations should be performed, and what risk monitoring should be put into place. For example, the risk of hallucination may be very important if a model is expected to be truthful (such as summarizing web search results), but perhaps irrelevant when truthfulness does not matter (such as brainstorming during a creative writing task). The model's use is the lens through which governance should be understood.

Thus, to effectively govern foundation models one needs an appropriate taxonomy of risks. This set of risks provides a palette that represents the possible risks relevant to a use of a foundation model. However, not all risks are relevant to all uses, so we also need a risk identification framework to identify which risks are relevant to a particular use. Once these risks are identified, one can test a candidate foundation model for those risks.
There has been significant work done on the first item (risk taxonomies) and the third item (benchmarks for testing for risks), but 
there has been relatively little
focus on the second item, leaving these two subfields disconnected.

Risk taxonomies focus on enumerating the various AI risks based on existing metrics and/or real-world AI incidents. Such taxonomies vary from "top ten" lists~\cite{owasp} to dozens~\cite{risk-atlas-paper, risk-atlas-url, ethics-board-pov}
or even thousands~\cite{mit-risk-repo} of risks. These taxonomies synthesize our current understanding of AI risks into a single place for quicker reference.

Concurrent with this line of research
is the development of effective benchmarks for testing for a particular risk (e.g., bias) or subrisk (e.g., gender assumptions in generated text) to determine how susceptible a model is to the risk. In addition to a robust dataset of prompts, this thread also needs a robust mechanism for scoring the model output, i.e., determining when model-generated text is biased with respect to a gender assumption. 

However, these research efforts are insufficient for identifying foundation model risks. More specifically, they do not typically account for how a foundation model is being used.
Although taxonomies adequately collect and define risks, they provide little to no guidance on how to apply the taxonomy to identify risks for a particular use. 
Benchmarks focus on testing models in ways that tend to be agnostic to how the model is used. For example, these benchmarks can provide a general sense of how fair or toxic a model's output might be, but those measures can be less meaningful when contextualized to a particular use. 
In short, neither taxonomies nor risk benchmarks serve the purpose of identifying \textit{which} risks are relevant to a use.

The goal of this paper is to provide guidance for developing a risk identification framework that addresses these gaps. Namely, how to identify the risks of AI foundation models based on use? How can we enable AI practitioners to understand the risks of their foundation model use cases and enable better foundation model governance? To answer these questions, we build on prior work in AI governance to determine the challenges in this space and how they can be addressed. Finally, we synthesize the solutions to the individual challenges and describe an initial design for a use-based AI risk identification framework.

We framed our risk identification framework through the lens of usage governance.
The key insight is that "use" is the primary determinant of possible risks.  Any risks that might be inherent in a model (e.g., hallucination), or where in a development process the risks might possibly arise, etc., must be interpreted within the context of that use.

The rest of this paper is organized as follows. 
Section~\ref{sec:related-work}
discusses related work in the domains of AI governance and AI risk assessment. 
Section~\ref{sec:usage-governance} describes
the necessary background for usage governance and identifies four design requirements for a risk identification framework.
Section~\ref{sec:design} considers
the implications of those design requirements in the context of risk identification.
Section~\ref{sec:risk-identification} describes how one might design a candidate risk identification framework to meet these design requirements.
Section~\ref{sec:value}
provides an illustrative example for risk identification and 
Section~\ref{sec:conc}
offers conclusions.

\section{Related Work} \label{sec:related-work}
AI risk taxonomies have been proposed by a number of organizations \cite{owasp, mitre, risk-atlas-url, risk-atlas-paper, mit-risk-repo, weidinger2022, uuk2024, abercrombie2024}. \citet{tax-to-reg} discuses the translation from taxonomy to regulation. Databases of harmful incidents involving AI provide concrete examples \cite{pai2020, incident2024}. \citet{incidents} describes a taxonomy for reporting incidents. Others have focused on AI risks that may be particularly salient within particular sectors including finance \cite{risks-finance, risks-finance-2}, law \cite{atkinson2024}, human relations \cite{risks-hr}, healthcare \cite{risks-health}, and defense \cite{risks-defense}. \citet{cross-domains} has discussed aspects of cross-domain taxonomies. Despite market pressures, there is no obvious way to identify any one of these taxonomies as universally correct or useful. Collectively, however, they do provide a rich space for consideration of possible AI risks.

A number of efforts have proposed templates and frameworks for describing AI dataset, model, and system properties \cite{datasheets-2018, datasheets-cacm-2020, dataset-nut-label-2gen-2020, data-statements, factsheets-2019, model-cards, aboutml-2020, shorensteincenter, derczynski2023assessing}. While risks are not the primary focus of these templates, they generally include at least a section on possible risks that could draw on the aforementioned taxonomies. 

Various efforts have focused on identifying and governing  AI risks. Some have a regulatory emphasis \cite{eu2024, canada2023, uk2023, china2025, nyc2021}. Others have a process emphasis \cite{nist-rmf, iso2023}.
Of these, the EU AI Act and NYC Local Law 144 begin to touch on aspects of the usage governance perspective we are exploring in the present work. The EU AI Act categorizes certain uses as being intrinsically higher or lower in risk (with various prohibitions and requirements depending on risk level). NYC Local Law 144 focuses on what must be done to assess, document, and audit bias for automatic systems involved in hiring, promotion, and similar work-related decisions.

Prescriptive aids and tooling for identifying AI risks are becoming available. 
The \citet{interparliament2025} has published sets of questions for identifying risks at each of the proposal, development, and operational phases of the AI life cycle.
The \citet{unicri2024} has published a questionnaire aimed at identifying risks for the use of AI within law enforcement. 
\citet{credoAI2025} provides a set of Policy Packs with associated online questionnaires to identify areas of non-compliance with one or more AI regulations.
\citet{herdel2024} discuss the use of large language models to assist in envisioning the uses and risks of AI.
\citet{wang2024farsight} describe and test a system that helps users identify potential AI risks during prompt-based prototyping.
For a broader perspective, \citet{koessler2023riskassessmentagicompanies} provides a useful overview of risk assessment techniques, including risk identification, for safety-critical industries.

Although these risk identification techniques provide a step towards helping an organization understand potential risks posed by an AI system, none of them produce a set of risks that reflect something like a fully contextualized use.
The goal of this paper is to understand what is needed for a risk identification framework to identify potential risks for a given foundation model use.

\section{Usage Governance and Design Requirements for Risk Identification} \label{sec:usage-governance}
This section describes usage governance and how it and other established challenges inherent to foundation model governance lead to the four key design requirements for a risk identification framework.

Usage Governance is a form of AI Governance that has emerged as a result of foundation models' increased capabilities.
The fundamental idea behind usage governance is that AI should be governed by how it is used. 
Before the introduction of foundation models, AI models tended to be created per problem (Should this loan be approved? What temperature should the A/C be set to? Does this image indicate a cancerous cell?). 
Data would be curated to specifically address each problem, and from the data each model would be trained. 
The association between the model and the problem being solved was strong so there was no need to distinguish governing the model from governing the problem. 
Model and use were effectively one and the same. 
Foundation models instead have more generalized problem solving capabilities. 
The same model can solve a multitude of problems, allowing for different uses for the same model. 
Consequently, the association between model and use weakened, resulting in a need to consider the use separately from the model. 

There are at least two advantages of decomposing risks in this way. First, it will highlight risks that might be otherwise missed. A model may, for example, have a relatively low rate of generating false summaries of documents. Whether this constitutes a concerning risk depends on how it is used; summarization of a movie review for personal use may be low risk while summarization of a legal contract may be high risk. Second, this decomposition will allow risk identification tooling to reuse at least some of the risk profiles of each entity without requiring them to be considered afresh with each change of context.

\subsection{Definitions}
Usage governance decomposes foundation model governance into several entities. Table~\ref{tab:definitions} provides a summary of these entities, and we detail them from a risk perspective further below.

\begin{table}[t]
    \centering
    \begin{tabular}{lp{2.25in}}
        \toprule
            Entity & Definition \\
        \midrule
            Use Case  & Domains of uses as often defined in regulations, ex: "hiring and promotion" \\
            Use       & The specific problem to be solved by the AI system \\
            Context   & The circumstances in which the model is deployed and used, including who is using it and what is output \\
            Data      & The data used to train or fine tune a model \\  
            Model     & The foundation model that is used \\
            Prompt    & The inputs, e.g., text, image, data set, given to the model to define its behavior either as an application prompt or part of a user's input \\
        \bottomrule
    \end{tabular}
    \caption{Usage governance entities and their definitions}
    \label{tab:definitions}
\end{table}

\subsubsection{Use Case}
A use case is a high-level construct that relates to a domain, such as ``hiring and promotion", ``education and vocational training", or ``law enforcement". Even with this high-level description, regulations exist to guard against certain risks.
For example, NYC Local Law 144 of 2021~\cite{nyc2021}
targets the use case of ``hiring and promotion" by requiring
a bias audit for any automated employment decision tools used for this use case. It also requires job candidates and employees to be notified of the use of such tools. Similarly, the EU AI Act~\cite{eu2024} imposes requirements for high-risk use cases, such as ``education and vocational training" and ``law enforcement".

\subsubsection{Use}
A use describes the specific problem that an AI solution is designed to address. It is more specific than a use case.
It answers the question
\emph{``What will the foundation model be used for?"}
For example, a use might be to use AI to analyze videos of applicant interviews. This brings specific risks beyond those identified in the general category of ``hiring and promotion".

AI researchers use the term \textit{task} to specify what a model needs to do, which may sound similar to a use.
However, these concepts are not the same.
The difference is that a use is a specific problem to be solved (Summarize a meeting transcript) while a task is the approach taken (document summarization). 

\subsubsection{Context}
The context provides additional details surrounding the circumstances of a particular use.
Context answers questions such as
\emph{``Who is interacting with the model?"}
\emph{``Who will be using the output generated by the model?"}
\emph{``Into what will the generated output be incorporated?"}
Use and context provide additional information that is needed to identify relevant risks. 
For example, if the analysis of interview videos is going to be distributed beyond the enterprise, additional risks are created.

\subsubsection{Training and Fine-Tuning Data}
The training data used to create a foundation model is usually not disclosed for scrutiny by others. Without this knowledge, any risks related to the data used to train the model, such as bias and falsehoods, can surface in the model's outputs. Additional legal concerns such as data ownership and privacy are more difficult to assess when training data are unavailable. Instead, evaluating such risks relies on prompting the model and detecting the risks across many outputs.

In some cases, additional model training is performed to fine-tune the model to a particular problem or domain. Unlike the data used to create the model, fine-tuning data are often under the direct control of the organization using the model because they are performing the tuning and can be more readily evaluated. 

\subsubsection{Model}
The model refers to the foundation model used to solve the problem.
The research evaluating models for risks in foundation models is extensive with measures to detect bias~\cite{parrish-etal-2022-bbq,tamkin-2023,kour2023unveiling}, adversarial robustness~\cite{bhatt2023purplellamacybersecevalsecure,zizzo-2024}, hallucinations~\cite{lin2022truthfulqameasuringmodelsmimic,chaudhury-etal-2022-x,mallen-etal-2023-trust}, 
and other types of risks~\cite{piorkowski-seton-hall}.

\subsubsection{Prompt}
Prompts are one of the primary ways to shape model behavior and control their outputs.
A prompt can refer to two different things. 
Most commonly, prompt refers to the input to the foundation model, usually in the form of text. 
The prompt is provided to the model, which provides some output (e.g. text, code, image, video, etc.) depending on the model. 
A prompt may also refer to instructions for how the model should behave. Sometimes referred to as a system prompt, this input may include a role the model should embody, specific behaviors to perform or not perform, instructions on how to format output, and the like. 
In practice, the input fed into the model is often a combination of the input prompt, the system prompt, and any other information that may help the model accomplish its goal (like past conversation history).

Prompts have their own specific risks.
Models have been repeatedly shown to be susceptible to a variety of prompt attacks used to manipulate the model's behavior~\cite{shi2024optimization, xullm2024}.
The effectiveness of these attacks have led to a variety of countermeasures and benchmarks that evaluate a model's robustness against such attacks~\cite{liu2024formalizing}.
Some of the factors that can contribute to the risk of a prompt include the text used in the prompt and the efficacy of any input sanitization techniques. 

\subsubsection{Design Requirements from Usage Governance}
Based on the identification of these six entities through the lens of usage governance, we can identify
two key design requirements for risk identification framework. First, if different entities carry different risks, then risks should be identified for each entity of the AI solution (including use and context as seen in Table~\ref{tab:definitions}). Second, if use decides how the model is governed, then use should decide which risks matter.  

In addition to these two design requirements for a risk identification framework, other established research inherent to foundation model governance provide two additional design requirements summarized below.

\subsubsection{Information About the Model Is Not Centralized} Prior work on how traditional AI developers build AI solutions has shown that AI system development is a complex, multi-stakeholder, multi-stage process. Some of the roles may include include domain experts, data labelers, data scientists, AI engineers, and external stakeholders. A typical AI model development process includes problem specification, data acquisition and transformations,  model development, application development that embeds the model, model and application validation/testing, application/model deployment, and monitoring. Furthermore, different roles enter and exit at those stages leaving a situation where information about how the system was built and the choices that were made are spread across different roles~\cite{piorkowski2021ai}.  It is unlikely that any one role has sufficient knowledge about the entire AI system's development.

Foundation models introduce additional complexity since they are often acquired or licensed for use instead of built from scratch. When a model is acquired, details of how it was created and evaluated for risk are often not provided. Available information about the model is often limited to documentation or white papers from the model provider and publicly available benchmarks evaluating the model. This provides additional complexity for foundation model risk identification since in the best case, the information to determine risks is spread over several places, and in the worst case, that information is not available. Additionally, foundation models introduce new sources of risk that model acquirers have to consider such as the prompt and fine-tuning data.

A risk identification framework should be designed for this reality. When information about risk is available, each entity's risks should be identified at the time that it is ready to be assessed by the people most familiar by that entity. This should reduce errors by involving the most knowledgeable people about an entity at a time where knowledge about a given entity is still available. To do so requires understanding the sources of risk and when should those risks be identified. Answering those questions leads to a two-part design requirement. The first part is a mapping from each risk to the entity or entities that contribute to that risk. The second part is a mapping from each risk to the stage in the foundation model development cycle (and its corresponding role) that the risk can be determined.

\subsubsection{Experience Is Limited} As noted earlier, the available benchmarks and risk taxonomies indicate that AI risk is a vast and quickly evolving field. With this pace of innovation, it is unlikely that any single member of an organization will know the details of each risk, their applicability, and how to mitigate it. The complexity in this space has the  potentially to overwhelm an end user interested in identifying the risks for their AI system. Addressing this complexity is at the crux of the final design requirement. Any risk identification framework needs to consider that its users may not know about the risks being identified, making user education as important as the main task of identifying risks.

\subsubsection{Summary of Design Requirements}
We now summarize the four design requirements mentioned in this section.
\begin{enumerate}
    \item \textbf{Entity Risks:} Risks should be considered across all entities of the AI system.
    \item \textbf{Use Contextualizes Risk} The selection of which risks are relevant should be defined by the use of the AI system.
    \item \textbf{Risk Mapping:} Risks need to be mapped both to the entities and stages of the AI development lifecycle that can identify that risk.
    \item \textbf{Usability:} Risk identification needs to understandable by end users who may have limited understanding about AI risks.

\end{enumerate}

\section{Implications for Design} \label{sec:design}
Although there are several candidate solutions to the design requirements noted above, here we detail the implications of each of the design requirements. 

\subsection{Entity-based Risks (Req. \#1)}
Since each entity can contribute different risks to the overall AI system, it is imperative to identify which risks each entity contributes. 
By framing risk as belonging to the entity, the possibility of reuse emerges.

Consider model risks such as hallucination and toxicity in output.
Once identified, the risks of this model stay inherent to this model.
Some of the risks may become more or less relevant as it is fine tuned, as guardrails are added, or as the way that the model's use changes, but the risks remain inherent to the model.
The same is true for other entities identified by usage governance.
We can illustrate this with some examples.
For context, if the model is deployed internally in an organization and controls are in place governing the use of the model, then prompt attack risks become less important.
For training data, if the data contains copyrighted material, then there is always a risk that copyrighted material will appear in the output of the model.
For prompt, if the prompt includes content from an end user, risks related to model misuse become more important.
Because the risks are tied to the entities, it may not be necessary to reidentify the risk of an entity when the entities around it change.
For example, if the same model is selected for a different use and a  different context, we would need to identify the risks for the new use and new context, but the risks of the model are already known.
This implication is useful given the reality that most organizations procure foundation models from third-parties.
Given that the cost of measuring and mitigating risks can be expensive, the savings for reusing a previous risk identification are notable.

\textbf{Implications}: A entity-based risk identification framework could enable reuse.

\subsection{Use Contextualizes Risk (Req. \#2)}
Throughout this paper, we provided varied examples that demonstrate how the use of a foundation model is the key to understanding the risks posed by an AI system. 
Although intuitive, defining the specifics of \textit{how} use contextualizes risk is harder to define. 
One reason is closely tied to the capability of foundation models. 
The extraordinary number of uses across multiple domains makes it difficult to distill which specifics matter. 
Some contextual information such as where the model is going to be deployed and who will use it can inform risks independent of the specifics of the use.
Yet, at least in the near term, we believe that the determination of what risks matter and to what degree is going to involve humans possibly informed by additional risk evaluation tools.

Assuming that this is a human process, the main implication is 
deciding what kind of information about the use and its context does a person need to determine if a risk is relevant. 
Candidate details include information about the use (What is the problem being solved? What information will be input to the  model? What task would the model perform?) and the surrounding context (Who are the users? Where is the solution going to be deployed? Is it internally or externally facing?).
Although these sorts of questions do not directly answer the question of if a risk is relevant, they help provide additional information.
That information is necessary to the person verifying if a risk is relevant and to be reasonably confident in doing so.

\textbf{Implication:} A risk identification framework must collect additional contextual information, in addition to what is needed to identify risks, to support those making the final decisions.

\subsection{Risk Mapping (Req. \#3)}

To understand entity risks, we need to be able to map risks to entities.
Fortunately, there is already some progress in this area. 
As mentioned earlier in the Usage Governance section, many benchmarks and taxonomies have already provided evidence of this mapping.
Benchmarks focused on risks are effectively measuring risks of models, implying that those risks map to the model entity.
However, some of those benchmarks are trying to identify some risks related to the data used to train the model, instead of the model itself such as those that aim to identify copyrighted data~\cite{ma2024inner}.
Regardless, it remains intuitive to map the risks from those benchmarks to the training data or other relevant entity.
Some taxonomies, such as IBM's AI Risk Atlas~\cite{risk-atlas-paper} already map risks to entities.
In that taxonomy, the risks are mapped to training-data risks, inference risks, output risks.
Together, benchmarks and taxonomies imply it is possible to map certain risks to entities.

Critically, there are going to be risks that do not map to the entities of the AI system itself. The taxonomies refer to these in different ways. 
In the MIT AI Risk Repository, these are identified as risks that do not belong to the AI entity. 
In the IBM Risk Atlas, these are found in the Non-Technical Risks topic. 
Many of these risks focus on governance, compliance, and societal risks and may not be easily identified from an AI system entity.
For example, impacts on jobs, communities, and the environment may require alternative approaches to identifying risks, that go beyond AI system entities.
For this paper, we do not consider these risks, and instead focus on those that can be be mapped to system entities.
However, we do not want to dismiss the importance of identifying these other risks.
We thus caution the reader that the approach in this paper should be considered as a part of a more comprehensive solution that would consider those additional risks.

The other mapping to consider is one from the risk to the stage of AI development.
The rationale is twofold.
The first is to understand at which stage of the AI development is there enough information to determine if a risk is relevant.
Some risks can only be identified if a entity exists (i.e., it does not make sense to identify model risks before a model has been selected).
However, other risks may be identified before the entity relevant to the risk is used or created.
For example, if the problem to be solved requires certain types of data, risks related to those types of data can be identified when a use is being defined.
Consider a use that requires an end user to provide personal information as input into the model.
Even before the model is selected and the prompt is formulated, we can infer that certain privacy risks will be relevant.
The second rationale is to identify what role can be responsible for identifying risks for a given stage.
Whether building a model from scratch or building atop a procured model, the knowledge gained about a entity tends to align with the stage of development.
People for one stage may not be the same as those in another.
For example, data scientists responsible fine-tuning the model may not be the same as those responsible for evaluating the model.
Clearly, if we are interested in learning about risks related to fine-tuning, we should rely on the data scientists who did the fine-tuning, that is, the people with the requisite knowledge should be responsible for identifying the risks for that stage or entity.

\textbf{Implications:} Risk identification should happen across different roles at different stages of the AI development lifecycle. Not all risks will map to a entity or stage of AI development.

\subsection{Usability (Req. \#4)}
As mentioned above, it is unlikely that the roles responsible for identifying risks will be knowledgeable enough about all the different kinds of AI risks. 
Thus, the user experience of a risk identification framework needs to be developed to support end users' decision making in the face of incomplete knowledge.
This support could take on many forms dependent on how risks are identified in the solution.
Prior work provides some possibilities.
The United Nations Interregional Crime and Justice Research Institute Risk Assessment Questionnaire provides examples of how risks may manifest~\cite{unicri2024}.
The Inter-Parliamentary Union Risk Assessment Questionnaire groups questions into categories or risk, allowing the user to focus in on the categories of interest~\cite{interparliament2025}.
A more comprehensive solution may include additional resources or a conversational agent informed about risks that an end user can interact with to understand a risks applicability.

Even with support, it is unlikely to have perfect certainty regarding the applicability of a given risk to an AI system.
To have such clarity, each risk would need to be considered by someone or a group with sufficient knowledge about that risk, an unlikely scenario given the vast number of risks.
Instead of a perfect identification of risks, we can instead aim to identify \textit{potential risks}.
We consider a potential risk to be one that is likely, given the available information, but still needs additional evaluation to be verified as relevant.
Although not ideal, identifying the subset of risks that should warrant further consideration is an improvement over doing so for all the possible risks.
In the context of the larger foundation model governance process, these risks would need additional assessment to determine their likelihood and consequentiality to determine if they are relevant or not.
That assessment could also incorporate benchmarks or other quantitative metrics of risk~\cite{piorkowski-seton-hall}.
Determining the relevance of a risk and its overall priority is a challenging problem and considered outside the scope of this paper.

\textbf{Implications:} A risk identification framework will need to support the end user's understanding of risk to be successful. Given the shortage of expertise, focusing on intermediate solutions like potential risks may be more appropriate.

\section{Candidate Solution} \label{sec:risk-identification}
This section describes how one might design a candidate framework for risk identification.
Our candidate solution is a set of questionnaires developed accordingly to the design requirements and implications above.
Recall that our goal is not to provide a complete risk identification framework, but to better understand what a solution needs to have.

\subsection{Solution Design}
We selected IBM's AI Risk Atlas~\cite{risk-atlas-paper} as the taxonomy for the candidate solution.
This process can start with any suitable taxonomy of AI risks, but we chose the AI Risk Atlas because it already is divided into risks about training data and model (inference and output risks in the atlas). 
This organization served as a good starting point for any solution as it provided a start to addressing design requirements \#1 (Entity Risks) and \#3 (Risk Mapping).
However, the risk-mapping requirement was incomplete.
It did not consider mapping to stages, nor did it map exactly to the entities defined in usage governance.

\subsubsection{Mapping Risks to Entities and Stages}

We started by selecting a small subset of the risks as candidates for further exploration, aiming to cover different risk categories and entities. 
We ended up selecting the risks of hallucination, toxic output, susceptibility to prompt injection attacks, model usage restrictions, and personal information in data.
To complete the mapping to the stages, we relied on existing benchmarks and a panel of AI risk researchers.
We started by identifying existing benchmarks for our selected risks. These benchmarks would inform what entity's risks were being currently evaluated and provided a starting point for what stage the risk could be considered.
For example, if a benchmarks was evaluated on a model, then we mapped the corresponding risk to stages involving model procurement or evaluation as appropriate.
However, several of the risks that we selected did not have available benchmarks to test those risks.
Consequently, we ended up relying on our panel of five AI risk researchers, one of whom is an author.
Each member of the expert panel had between five and nine years of professional experience researching and implementing solutions to detect and mitigate AI risks.

We asked our AI researcher panel to review both the risks that we mapped to stages of AI development from the benchmarks and the ones we were unable to map.
We specifically asked the panel for two kinds of information.
The first ask was their expert opinion on what stage or stages each risk could be considered. The second ask was the conditions that if true, suggested that a risk should be considered a potential risk.

The expert panel mapped all five of our selected AI Risk Atlas risks to one or more entities from Usage Governance and one or more stages of foundation model development. The risks that we selected ended up mapping to three different stages. The first is the use definition stage, where the problem to be solved is defined. The second was the model procurement stage, where a model is acquired by an organization. The third was the implementation stage, where a model is purposed for a use. 
There are additional possible stages such as evaluation, deployment and monitoring. 
Including other risks may have resulted in additional stages being considered, but the necessity of having a questionnaire at different stages was already clear.

The expert panel also provided the conditions under which should a risk be considered potentially relevant.
For those five risks, the expert panel detailed a total of fourteen different conditions.
The number of conditions per risk varied from one to five per risk.
For example, the hallucination risk included, ``If falsehoods or misinformation were found and not removed from the training data, then falsehoods or misinformation may appear in model output,'' and ``If the model’s input includes content created by people, then the input may include instructions that cause the model to generate falsehoods or misinformation in its output'' among its five conditions.

We then converted each condition to a question or set of questions that matched the condition. For example, the second condition above was checked using the question, ``Will the model input include content provided or created by people?'' and ``Are you planning to use generative AI model(s) to address this use?'' The first question confirms that input may have human-generated content, and the second question confirms that we are expecting to use a generative model. Together, these meet the condition expressed by our experts.

Finally, we determined what roles would be appropriate to answer at the three stages relevant to our risks. For the first set of risks, those that mapped to the use definition stage of development, we realized that the role answering these questions likely includes people without a data science background. Product owners, domain experts, and general users who want to use a foundation model to accelerate their work are all candidates who may be answering this questionnaire. The other two stages, model procurement and implementation will require data scientists. These risks require understanding foundation model documentation and benchmarks and thus, require the corresponding data science background to answer the questions.

The end result was a set of three questionnaires, one for each of the three stages, deemed relevant by the five risks we considered.

\subsubsection{Addressing Usability}

In an effort to address design requirement \#4 (Usability), we incorporated additional information that the questionnaire answerer can use to inform their answer. 
This text provided additional explanations to provide more clarity to the question, or it provided guidance to help the answerer consider aspects of the question they may otherwise not consider.
For example, the additional text to the question about if model input contains content created by people, the additional text was

\begin{quote}
    \textit{Inputs to non-generative AI models may include user content if they come from an interface that an end user interacts with. This may occur in real-time or if the information is stored and then used by the model later.}
    
    \textit{For generative AI models, model input may contain a system prompt (a set of instructions provided to the model), user prompt (content provided by end users), and additional context (any additional input provided to improve model output). If any of these contain end-user input, that is considered as input to the model. As an example, a system prompt is a set of instructions provided to the model. For example, “You are a helpful, friendly, …” This is different from an input or user prompt, which is the text sent to the the model as input.}
\end{quote}
We provided additional text such as this throughout the questionnaires.

\subsubsection{Contextualizing Risk}

Although the questionnaire is capable of identifying potential risks, it does not help determine which risks are relevant. To do so, context is needed. Throughout the questionnaire we included open-ended questions to build this context. These questions were designed to distinguish one use from another. This information could then be used by the person or people responsible for determining if a particular risk was relevant.

These questions fall into two categories. The first category is derived from the definition of context from usage governance. These could include questions such as those about the model's owner's users, and affected persons. They can also include questions about the model, such as where it was deployed, how it was going to be used or could be misused. The second category of questions are information collection questions. Since model risks are likely to come from existing evaluations noted in model documentation or other benchmarks of these models, recording what was measured, how it was measured, and the results of those measurements provide necessary context to understand model risks. Examples of both types are provided in the next section.

\section{Illustrative Example} \label{sec:value}

\begin{table*}[]

    \small
    \begin{tabular}{p{0.2in}p{1.8in}p{4.5in}}
    \multicolumn{3}{c}{\textbf{Use Questionnaire}} \\
    \toprule
    \#  & Question & Additional Description \\
    \midrule
    A1. & Please describe the problem you are solving with AI. & n/a \\
    A2. & Please describe the expected users of the model. & The expected users of the model are those who use the model’s output. \\
    A3. & Please describe the persons expected to be affected by the model. & The persons affected by the model are the people who may be impacted by a model’s output, even if they never interact with the model themselves. \\
    A4. & If the model is part of a larger solution, please describe its role in the larger solution. & Consider whether this model is just one piece of a solution to a larger problem. For example, a model that predicts the likelihood of someone repaying a loan could be part of a larger solution that determines if a loan should be approved. When answering this question, consider the aspects of the larger solution that someone will need to know to evaluate risks associated with this use case. \\
    A5. & What types of data are expected to be used as input? & A use may require multiple types of data to determine an answer. For example, a loan determination may take into account the size of the loan, the applicant’s income, and other factors to determine if a loan should be approved. These data points may be used as inputs for an AI model. For generative models specifically, input often takes the form of text-based prompts that include the required data. \\
    A6. & Will input include content provided or created by people? & Inputs to non-generative AI models may include user content if they come from an interface that an end user interacts with. This may occur in real-time or if the information is stored and then used by the model later. For generative AI models, model input may contain a system prompt (a set of instructions provided to the model), user prompt (content provided by end users), and additional context (any additional input provided to improve model output). If any of these contain end-user input, that is considered as input to the model. As an example, a system prompt is a set of instructions provided to the model. For example, “You are a helpful, friendly, …” This is different from an input or user prompt which is the text sent to the the model as input. \\
    A7. & Could users of this model include malicious users outside your company? & Malicious users are people who want to benefit from the model in an unintended way or cause harm. Consider if the use case might have malicious users outside the company who interact with the model. \\
    A8. & Would those malicious users be able to send inputs to the model and see its output? & If malicious users outside the company can supply inputs to the model and also view its outputs, they could deduce information that makes model attacks more feasible.  \\ \bottomrule
    \\ \\ 
    \multicolumn{3}{c}{\textbf{Model Onboarding Questionnaire}} \\
    \toprule
    \#  & Question & Additional Description \\
    \midrule
    B1. & Was the model's training data screened for hateful, abusive, or aggressive content? & Examples of hateful, abusive, or aggressive content include inappropriate language or imagery, hate speech, and discriminatory or derogatory terms. \\
    B2. & Was hateful, abusive, or aggressive content found in the training data? & n/a \\
    B3. & Was that content removed from the training data? & n/a \\
    B4. & Summarize how that content was removed, or provide a link to evidence. & n/a \\ \bottomrule
    \\ \\ 
    \multicolumn{3}{c}{\textbf{Use and Model Questionnaire}} \\
    \toprule
    \#  & Question & Additional Description \\
    \midrule
    C1. & Does the models terms of use permit this use? & Models may have usage or licensing terms that specify how they can be used. Some terms of use explicitly prohibit the use of models for certain use cases. \\
    \bottomrule
    \end{tabular}
    \caption{The complete example questionnaires with all descriptive text}
    \label{tab:questionnaire-full}
\end{table*}

In this section we present a partial solution as an illustrative example.
We only consider three of the five foundation model risks in our example to demonstrate how the different concerns manifest, as the two risks we are not presenting had a very similar structure to the ones we are, and provide no additional insights.
The risks we present are toxic output, susceptibility to prompt injection attacks, and model usage rights restrictions. 
Our goal here is demonstrate how usage governance and the design guidelines shape the development of a risk identification framework.
Table~\ref{tab:questionnaire-full} details the three questionnaires.

This example aims to identify potential risks and assumes that those risks will be then re-reviewed to determine how relevant they are.
Thus, the questionnaire consists of questions for identifying the potential risks and also additional contextual questions aimed at providing the reviewers the relevant content necessary to make an informed decision.

For this example, consider a company that wants to develop a question answering conversational agent set up in a visitor center at a local tourist attraction. 
To identify the potential risks of this AI system, the project owner would start by filling out the Use Questionnaire. 
Questions A1 to A5 collect necessary context about the problem being solved. 
They are not used to identify the potential risks directly, but instead are meant to gather information for later use when a risk is considered as relevant.
Question A6 to A8 capture information that is needed to determine if toxic output or prompt injection attacks are a potential risk.
We will detail how the question responses lead to a potential risks later in the section.

Once a model is selected to implement the use, a data scientist fills out the Model Onboarding Questionnaire.
Unlike the prior questionnaire, which required no external sources, this questionnaire requires the data scientist to review available documentation and public benchmarks for the acquired model to answer the questions.
Unlike the Use Questionnaire, the answers to these questions may not be available as foundation models tend to not be very transparent.
Thus, a valid answer to questions B1 to B4 can be "Not found in the documentation."
Like before, questions B1 and B2 capture information to determine potential risk and questions B3 and B4 summarize information to help determine if the risk is relevant.

The last questionnaire can only be answered once the use is defined and the model is selected.
In this case, the question directly answers the risk of model usage rights restrictions, asking the person to verify that what the use does is within the model's terms' allowed uses.

Recall that we based our questions on an AI risk researcher panel's conditions. 
Thus, different combinations of responses will either suggest or not suggest a potential risk. 
Considering the risk of toxic output, one condition provided by our experts is ``If the training data was not screened for toxic content, then the model may generate inappropriate language or imagery, hate speech, or discriminatory and
derogatory terms.''
Responses to questions B1 to B3 check if training data was screened, if the screening found toxicity, and if it was addressed. 
Clearly, if toxic content was found and not suitably addressed, the risk of toxic content remains a possibility, and it is flagged as a potential risk.
But what if there is no mention of a toxicity assessment in the model's documentation and no benchmarks results for toxicity for this model either?
In that case, we believe that the risk should also be flagged as a potential risks because there is no evidence to the contrary.
However, this is ultimately a decision for the organization's governing body and their tolerance for risk. 

The risk of prompt injection attacks is handled similarly. In this case, the condition was ``If a malicious user can prompt the model, then they may force the model to
produce unexpected output by manipulating their prompt.'' This condition was converted to questions A6 to A8. 
For this use, since people will be interacting with the question answering agent, can observe output for a given input, and have the potential to be malicious, then the susceptibility to prompt injection attacks is flagged as a potential risk.
Recall that once all the questionnaires are filled out, the list of potential risks undergo additional review to determine if they are relevant to the use.
These reviewers benefit from the information collected about the problem being solved (A1) and expected users of the AI system (A2) in making their determination of each risks' relevance.

\section{Conclusions} \label{sec:conc}

Although current foundation model risk identification techniques are structured to help people think about risks in AI systems, they stop short of recommending specific risks contextualized to a use.
To address this gap, we extracted design requirements to address this need from usage governance and prior literature.
We identified the following four design requirements:
\begin{enumerate}
    \item \textbf{Entity Risks:} Risks should be considered across all entities of the AI system.
    \item \textbf{Use Contextualizes Risk:} The selection of which risks are relevant should be defined by the use of the AI system.
    \item \textbf{Risk Mapping:} Risks need to be mapped both to the entities and stages of the AI development lifecycle that can identify that risk.
    \item \textbf{Usability:} Risk identification needs to understandable by end users who may have limited understanding about AI risks.

\end{enumerate}

We considered the implications of these design requirements in the context of developing a risk identification framework. 
Then, using those design guidelines, we proposed a candidate solution in the form of three questionnaires targeted at different roles and at different stages of implementing a foundation model use.
Finally, we explained in detail how the questionnaire was created and how responses to questions identified potential risks via an illustrative example.
Our goal in this formative work 
is to accelerate the development of better frameworks for identifying foundation model risks.

\bibliographystyle{ACM-Reference-Format} 
\bibliography{refs}

\end{document}